\documentclass[aps,prl,twocolumn,superscriptaddress]{revtex4}
\usepackage{bm}
\usepackage{graphicx}
\usepackage{amssymb,amsmath,amsbsy,amsgen,amsfonts}
\usepackage{dcolumn}
\usepackage{amsthm}
\usepackage{mathrsfs}
\usepackage{latexsym}
\usepackage{array}
\usepackage{amstext}
\usepackage{epsfig}
\usepackage{epstopdf} 

\usepackage{color}
\usepackage{units}
\usepackage{calrsfs}
\DeclareMathAlphabet\mathbfcal{OMS}{cmsy}{b}{n}

\newcommand{\be}{\begin{equation}}
	\newcommand{\ee}{\end{equation}}
\newcommand{\ba}{\begin{array}}
	\newcommand{\ea}{\end{array}}
\newcommand{\bqa}{\begin{eqnarray}}
	\newcommand{\eqa}{\end{eqnarray}}

\begin{document}

\title{Topological Wave-Guiding Near an Exceptional Point: \\ Defect-Immune, Slow-Light, Loss-Immune Propagation}

\author{S. Ali Hassani Gangaraj}
\address{School of Electrical and Computer Engineering, Cornell University, Ithaca, NY 14853, USA}

\author{Francesco Monticone} \email{francesco.monticone@cornell.edu}
\address{School of Electrical and Computer Engineering, Cornell University, Ithaca, NY 14853, USA}

\date{\today}

\begin{abstract}

Electromagnetic waves propagating, at finite speeds, in conventional wave-guiding structures are reflected by discontinuities and decay in lossy regions. In this Letter, we drastically modify this typical guided-wave behavior by combining concepts from non-Hermitian physics and topological photonics. To this aim, we theoretically study, for the first time, the possibility of realizing an exceptional point between \emph{coupled topological modes in a non-Hermitian non-reciprocal waveguide}. Our proposed system is composed of oppositely-biased gyrotropic materials (e.g., biased plasmas or graphene layers) with a balanced loss/gain distribution. To study this complex wave-guiding problem, we put forward an exact analysis based on classical Green's function theory, and we illustrate the behavior of coupled topological modes and the nature of their non-Hermitian degeneracies. We find that, by operating near an exceptional point, we can realize anomalous topological wave propagation with, at the same time, low group-velocity, inherent immunity to back-scattering at discontinuities, and immunity to losses
. These theoretical findings may open exciting research directions and stimulate further investigations of non-Hermitian topological waveguides to realize robust wave propagation in practical scenarios.

\end{abstract}

\maketitle

\emph{Introduction} --- The ability of guiding light and electromagnetic waves in desired directions is of fundamental importance for science and technology, from long-haul optical fibers, to the possibility of on-chip integrated optical interconnects \cite{Miller}, and enhanced light-matter interactions at the micro- and nano-scale \cite{Novotny}.

Wave propagation in conventional wave-guiding structures is characterized by some typical properties that limit their behavior and applicability: (i) Conventional waveguides support modes that are allowed to propagate in either directions with the same (opposite) wavevector, $|k_{+}|=|k_{-}|$, i.e., they are symmetric upon time reversal, a consequence of the Lorentz reciprocity theorem for most media and structures \cite{Jackson}. As a result, any defect and discontinuity in the waveguide is allowed to excite a backward-propagating wave. (ii) Typical waveguide modes exhibit moderately large group/energy velocity, $\partial \omega / \partial k$, hence limited interaction-time with matter. Specific dispersion-engineering strategies are necessary to slow light down \cite{Khurgin}. (iii) Waveguide modes decay in lossy regions, which obviously limits their propagation length. More generally, it is typically accepted that, in transversely-homogeneous waveguides, the amplitude of a guided mode, $\propto |e^{-i \omega t} e^{i \textbf{k} \cdot \textbf{r}}|$, with respect to the direction of propagation, is either a constant function if the waveguide is closed and lossless, or an exponential function if the waveguide is lossy/gainy. 

In this Letter, we theoretically propose a wave-guiding structure that violates all three points outlined in the previous paragraph at the same time, based on combining concepts from non-Hermitian physics and topological photonics. This may lead to the realization of anomalous guided-wave propagation exhibiting backscattering-immunity, loss-immunity, and low group velocity simultaneously. To realize this exciting possibility, we present for the first time an \emph{exact} analysis of coupled unidirectional modes in non-Hermitian topological wave-guiding structures based on Green's function theory, which reveals the presence and nature of an exceptional point (EP) where two topologically-protected modes coalesce. Such a topological non-Hermitian degeneracy is the key to the anomalous propagation properties described above.



\emph{Topological non-Hermitian wave-guiding} --- Our general discussion is based on a continuum model of a photonic topological insulator with a broken time-reversal symmetry (a Chern-type insulator) \cite{Chern_type_1,Chern_type_2}, but our considerations can be extended to any type of photonic topological materials \cite{Rechtsman1,Zhang_PRL,Soljacic_PRL}. In particular, we consider a homogeneous or effective gyrotropic material, which can be described by permittivity and permeability tensors: $ \underline{\boldsymbol{\epsilon}} = \epsilon_0 \left( \epsilon_{t} \boldsymbol{\mathrm{I}}_t + \epsilon_{a} \hat{\boldsymbol{\mathrm{z}}} \hat{\boldsymbol{\mathrm{z}}} + i \epsilon_{g} \hat{\boldsymbol{\mathrm{z}}} \times \boldsymbol{\mathrm{I}}   \right) $, $ \underline{\boldsymbol{\mu}} = \mu_0 \boldsymbol{\mathrm{I}} $ such that ${\boldsymbol{\mathrm{I}}}_{t}=\boldsymbol{\mathrm{I}}-\mathbf{{\hat{z}}{\hat{z}}}$ ($ \boldsymbol{\mathrm{I}}  $ is the unit tensor), with $\varepsilon_g$ being the magnitude of the gyration pseudovector. A continuous gyrotropic material can be realized, for example, by a plasma magnetically biased along a given direction (in our case, the $ z $-axis) with frequency-dispersive permittivity elements, having a certain plasma frequency $\omega_p$ and cyclotron frequency $\omega_c$ \cite{Bittencourt,SM}.
%
It has been known since at least the 1960s that, when a gyrotropic material is interfaced with a different medium, under certain conditions a unidirectional scattering-immune surface mode emerges \cite{1962}. In recent years, the nature and origin of this mode has been related to the non-trivial topological properties of the gyrotropic material \cite{Mario_1,Hassani_1,Hassani_2,Hassani_3}. In particular, for the case of a magnetized plasma, the gap Chern number is equal to unity \cite{Hassani_1}; therefore, at the interface with a topologically-trivial material exactly one topologically protected one-way edge state emerges.

\begin{figure}[t]
	\begin{center}
		\noindent \includegraphics[width=0.48\textwidth]{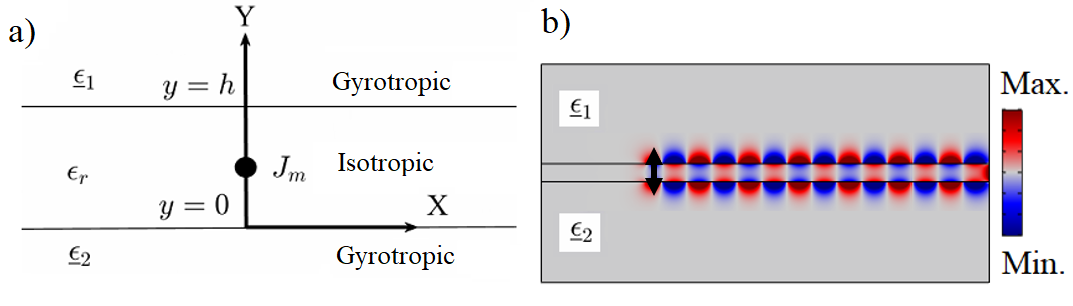}
	\end{center}
	\caption{(a) Topological wave-guiding structure, composed of two oppositely-biased gyrotropic media, separated by an isotropic layer ($J_m$ indicates a current source). 
    (b) Example of field distribution (time-snapshot) propagating in a lossless (Hermitian) topological waveguide as in (a), excited by a dipolar source (black arrow). Two unidirectional surface waves propagate along the interfaces and couple through the isotropic layer. For this example, we assumed $ \omega_{p,1} = \omega_{p,2}  = 0.9 \omega_0  $, $ \epsilon_r = -1 $, and $  \omega_{c,1} = - \omega_{c,2}  = 0.28 \omega_0   $, where $ \omega_0 $ is the source frequency.}
	\label{structure}
\end{figure}


In order to study a configuration with coupled topological modes, we then consider a layered wave-guiding structure as in Fig. \ref{structure}(a), composed of a topologically-trivial isotropic layer of thickness $h$ and permittivity $ \epsilon_r $ sandwiched between two gyrotropic media with permittivity tensors $ \underline{\boldsymbol{\epsilon}}_i $, $ i=1,2 $, which we name TNT (Topological Non-topological Topological) waveguide. The two gyrotropic media, for example magnetized plasmas, are oppositely biased in order to have unidirectional transverse-magnetic surface waves propagating in the same direction, as shown in Fig. \ref{structure}(b). The two surface waves are topologically-protected and can couple if the two interfaces are sufficiently close.

The exact Green's function of the structure can be expressed as a Fourier integral with respect to the transverse momentum $ k_{x} $ (further details in \cite{SM}). In the central region of the TNT waveguide, assuming the frequency of operation lies within the bandgap of the biased plasmas, this integral can be evaluated as a sum of residue terms, corresponding to the unidirectional guided modes of the waveguide, as follows: 
\begin{equation}\label{Res}
H_z(x,y) = 2 \pi i \sum_m  w^{\mathrm{spp}} (k_{mx}, \omega) e^{ik_{mx}x},
\end{equation}
where $ w^{\mathrm{spp}}(k_x, \omega) = N(k_x, \omega)/\partial_{k_x} D(k_x, \omega) $ is the complex amplitude of the mode, and
%
%
the functions $N$ and $D$ are defined in \cite{SM}. $ k_{mx} $ is the $m$th root ($m$th guided-wave pole) of the dispersion equation of the system $ D(k_x, \omega) = 0 $, for a given set of parameters $(\omega, h, \omega_{pi}, \omega_{ci}) $. If the system is closed (Hermitian), with no decaying channels (intrinsic loss or radiation), the modes of the waveguide are orthogonal \cite{VanBladel}, whereas, if the system is open, two (or more) modes may perfectly coalesce and become linearly dependent at an EP \cite{Berry_EP,Heiss_EP,Rotter_EP,Capolino,Christodoulides}, leading to interesting propagation effects near the degeneracy.



\begin{figure}[b]
	\begin{center}
		\noindent \includegraphics[width=0.495\textwidth]{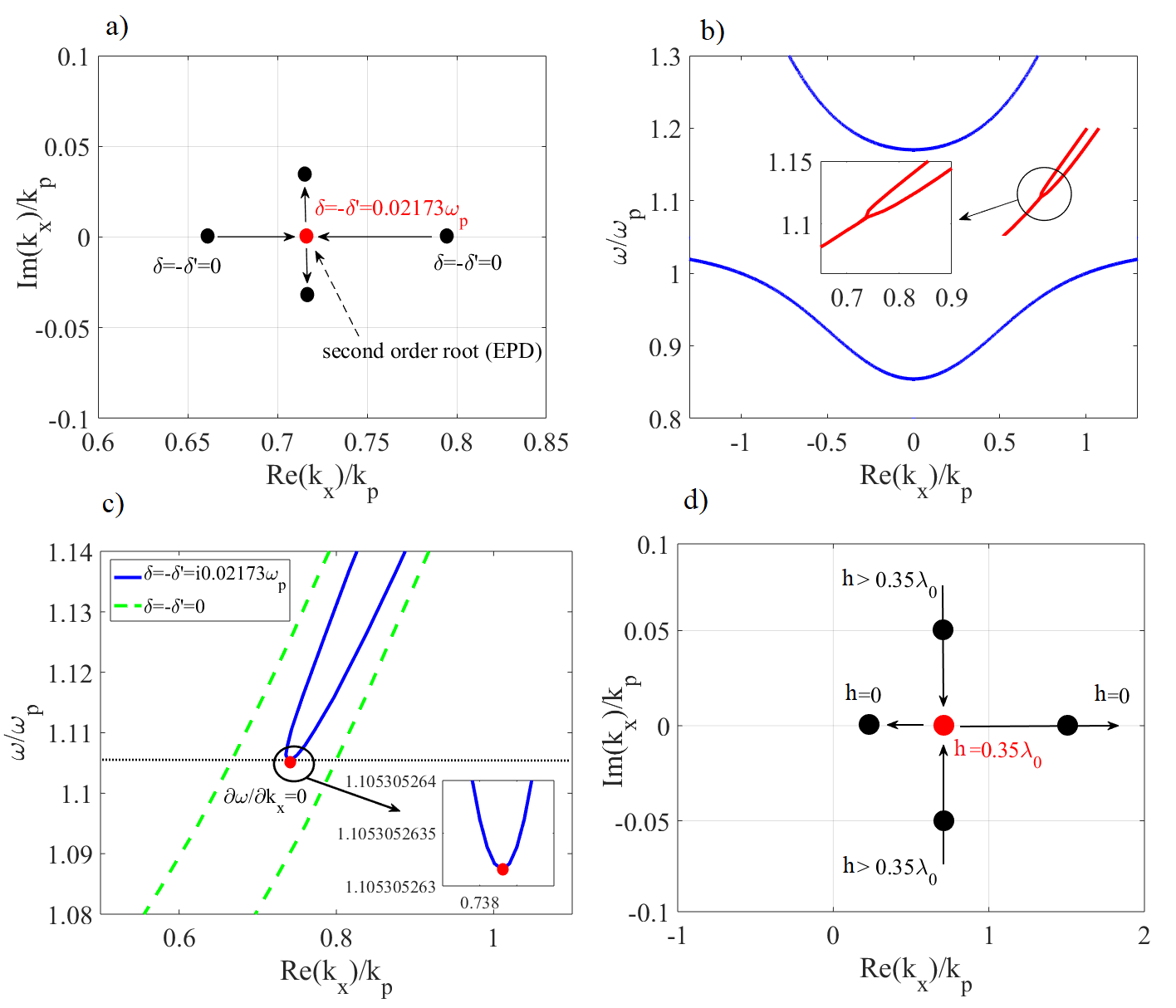}
	\end{center}
	\caption{(a) Evolution of the guided-wave poles in the complex wavenumber plane for the topological waveguide in Fig. 1, as a function of the level of loss/gain in the system (with $ \omega_p/\omega_0 = 0.904 $, $ \omega_c/\omega_0 = 0.287  $ and $ h = 0.35 \lambda_0 $). Black dots indicate the roots of the dispersion equation, and the red dot denotes the EP. (b) Dispersion curves of the bulk (blue) and topological modes (red) for a TNT waveguide with $ \omega_p/\omega_c = 3.16 $, $ \delta = - \delta' = \delta_{EP}$, $\omega_p $, $ h = 0.63 \pi c/\omega_p $ and $ \epsilon_r = -1 $. (d) Comparison between the dispersion curves of the topological unidirectional modes in a Hermitian system (dashed green) and a non-Hermitian system (blue). (d) Similar to (a) but varying the isotropic gap thickness (with $ \delta = -\delta' = \delta_{EP} $).}.
	\label{fig3}
\end{figure}

To study the non-Hermitian topological case, we introduce loss and gain into the gyrotropic layers of the TNT waveguide. Mathematically, this is obtained by modifying the frequency dependent elements of the permittivity tensors as $ \epsilon_{1,j}(\omega + i\delta, \omega_p, \omega_c) $ and $\epsilon_{2,j}(\omega - i\delta', \omega_p, \omega_c)$, where $ ~ j = t,a,g$. By choosing $ \delta = - \delta' $, we realize a balanced (parity-time symmetric) configuration. The trajectory of the guided-wave poles, in the complex wavenumber plane, for this wave-guiding configuration is shown in Fig. \ref{fig3}(a), as the level of loss/gain is increased. When $  \delta = -\delta'=0 $, we have two poles on the real axis with $ \mathrm{Im}(k_x) = 0 $, corresponding to the Hermitian case. Due to the nonreciprocal nature of the structure, there are no symmetric poles on the negative side of the real axis. Then, as we increase the level of loss/gain, the two poles move closer to each other until, for $ \delta = - \delta' = \delta_{EP}= 0.02173 \omega_p $, the poles collide and merge at $ k_x = 0.738 \omega_p /c_0 $ (red dot in the figure). 
Different from most previous works on non-Hermitian photonics, our exact analysis based on Green's function theory and continuum material models allows us to rigorously verify that this point of the parameter space is indeed an EP between topological modes. A necessary condition is that two first-order roots of the dispersion function $ D(k_x, \omega) $ coalesce to form a second-order root, which implies that the dispersion equation at an EP should satisfy: 
\begin{equation}\label{1st_cindition}
D(k_{x,\mathrm{EP}}, \omega_{\mathrm{EP}}) = \frac{\partial D(k_{x,\mathrm{EP}}, \omega_{\mathrm{EP}})}{ d k_x} = 0,
\end{equation}
where $ k_{x,\mathrm{EP}}$ and $  \omega_{\mathrm{EP}}  $ are the wavenumber and angular frequency at which the degeneracy emerges. However, the condition in (\ref{1st_cindition}) guarantees nothing more than the existence of a double root, whereas an EP should also satisfy the additional condition 
\begin{equation}\label{2nd_cindition}
\frac{\partial D(k_{x,\mathrm{EP}}, \omega_{\mathrm{EP}})}{ d \omega} \cdot \frac{\partial^2 D(k_{x,\mathrm{EP}}, \omega_{\mathrm{EP}})}{ d k_x^2}  \neq 0.
\end{equation}
These two conditions are necessary and sufficient to guarantee that the pair $ (  k_{x,\mathrm{EP}}, \omega_{\mathrm{EP}}) $ corresponds to a true EP \cite{RS_1998,RS_1999,Hanson_TAP_2003}. Indeed, only if (\ref{2nd_cindition}) is satisfied, then the second-order root becomes a branch point where various branches of the dispersion function merge \cite{RS_1998,RS_1999} (and not a saddle point of the dispersion curve). Numerical tests applied to our specific case verify that our dispersion function near the red point in Fig. \ref{fig3}(a), with $\omega_{\mathrm{EP}} = \omega_p/0.904 $, satisfies conditions (\ref{1st_cindition}) and (\ref{2nd_cindition}), hence fully confirming the existence of a branch point where two distinct \emph{topological} modes coalesce. Moreover, by expanding a generic dispersion equation $ D(k_x, \omega)$ in a power series near $k_{\mathrm{EP}}$ and $\omega_{\mathrm{EP}}$, it is easy to show that conditions (\ref{1st_cindition}) and (\ref{2nd_cindition}) are satisfied if the local structure of the dispersion function has the form $ D(k_x, \omega) = a (k_x - k_{\mathrm{EP}})^2  \pm b (\omega - \omega_{\mathrm{EP}} )$, where $a$ and $b$ are generic complex numbers. A first-order EP is therefore identified as a square-root branch point \cite{Seyranian,Moiseyev,MTT-2000}. Note that, in our case, the dispersion function is not symmetric in $k_x$ due to the nonreciprocal nature of the TNT waveguide. Such a nonreciprocal square-root-like structure of $ D(k_x, \omega)$ has crucial implications for the propagation properties of the proposed structure, as discussed in the following.

As we increase the level of loss/gain in the system above the critical value $\delta_{EP}$, the EP breaks into two complex conjugate poles (Fig. \ref{fig3}(a)), leading to exponentially decaying/growing unidirectional modes propagating along the $+x$-axis. This is a typical behavior for parity-time-symmetric systems, 
but with the important difference that in our case the modes involved in this modal (phase) transition are topologically-protected and back-scattering immune. 
 


\emph{Topologically-protected slow-light modes} --- The dispersion curves of the coupled topological modes are shown in Fig. \ref{fig3}(b), where the loss/gain level is set at the critical value $\delta_{EP}$
, clearly demonstrating the \emph{bifurcation} of the topological modes (red lines) and the emergence of an EP within the band gap of the bulk modes (blue lines) \cite{note1}. For frequencies $ \omega > \omega_{\mathrm{\mathrm{EP}}} $, there are two modes with $ \mathrm{Im}(k_x) = 0 $, whereas for $ \omega < \omega_{\mathrm{EP}} $ the modes depart from the wavenumber real axis and form two complex conjugate modes with equal $ \mathrm{Re}(k_x) $. In the neighborhood of the EP, the dispersion behavior is locally determined by $ k_x - k_{\mathrm{EP}}  \propto \sqrt{ \omega - \omega_{\mathrm{EP}} } $. This relation directly implies that the dispersion curves flatten out before merging, such that $ \partial_{\omega} k_x \rightarrow \infty $; hence the group velocity vanishes, $ v_g \rightarrow 0 $. Fig. \ref{fig3}(c) compares the dispersion of the closed system with no loss/gain and the open system with equal amounts of loss and gain. This confirms that a waveguide working at the EP may be used to slow down and stop light, as recently observed in \cite{light_stop}. However, in drastic contrast with any other light-stopping system proposed so far, in our TNT waveguide \emph{the slowing down of light is topologically protected}, and tunable by changing the bias. These features may suggest novel functionalities, in which the propagation of an electromagnetic wave can be fully stopped, and then released, by varying the parameters of the system, e.g., the thickness of the waveguide, or the level of gain/loss, or the external bias, without losing any energy in unwanted back-scattering, thanks to the waveguide's topological properties.

\begin{figure*}[bth]
	\begin{center}
		\noindent \includegraphics[width=0.85\textwidth]{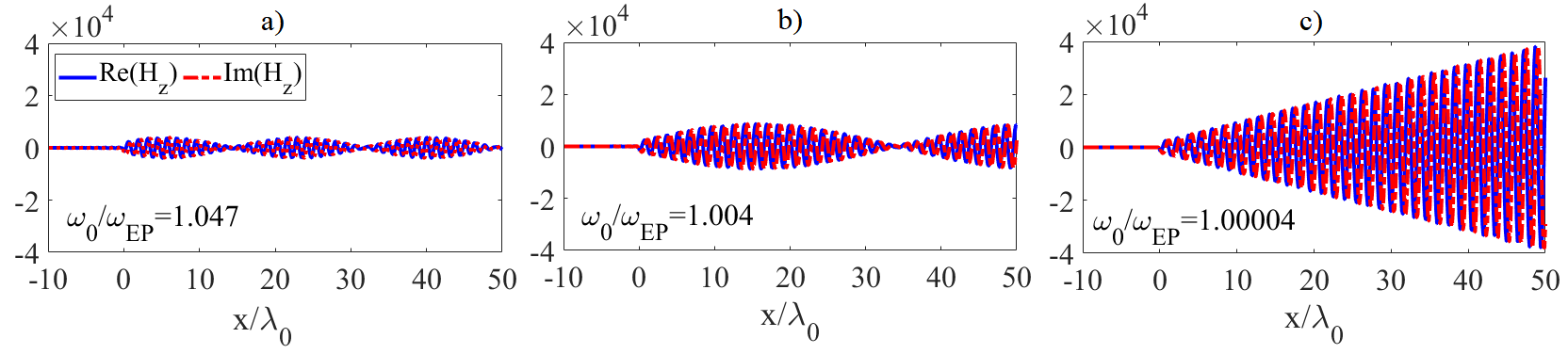}
	\end{center}
	\caption{Magnetic field distribution at the center of the topological waveguide considered in Fig. 2(b), for a dipolar source at $x=0$, as its frequency $\omega_0$ approaches $ \omega_{\mathrm{EP}}= \omega_p/0.904 $ (exact Green's function calculations).}
	\label{fig4}
\end{figure*}

\emph{Linearly-growing topological modes} --- Due to its nature as a second-order singularity of the Green's function, as required by (\ref{1st_cindition}), an EP leads to divergence of the residue terms $ w^{\mathrm{spp}}(k_x, \omega) = N(k_x, \omega)/\partial_{k_x} D(k_x, \omega) $ in Eq. (\ref{Res}), associated with the two topological modes on the verge of merging. As a result, the field intensity tends to grow larger and larger as we operate closer to the EP. While no realistic structure would be able to exactly operate at the EP (any, unavoidable, asymmetry in the gain/loss distribution would move the EP off the real axis), it is relevant to study the evolution of the unidirectional fields as we approach the degeneracy. Fig. \ref{fig4} shows the magnetic field distribution in the TNT waveguide, calculated exactly based on our Green's function formulation, for a dipolar source at $x=0$ at a frequency near the EP. The interference of the two topological waveguide modes forms unidirectional ``wave packets'', which become longer in length and larger in intensity as we get closer to the EP (Fig. \ref{fig4} (a-c)). In particular, as we approach the frequency of the degeneracy, and the difference in wavenumber between the modes gets smaller $\Delta k_x = k_{x,1} - k_{x,2} \rightarrow 0$, the residue terms tend to acquire a phase difference of $\pi$ \cite{SM}, and we can write the field distribution by Taylor expanding Eq. (\ref{Res}), obtaining 
\begin{equation}\label{Taylor}
	{\mathop{\rm Re}\nolimits} \left[ {{H_z}} \right] \approx \mathcal{D}(\Delta k_x) \frac{{\Delta k_x x}}{2} \sin \left[ {\frac{{\left( {{k_{x,1}} + {k_{x,2}}} \right)x}}{2}} \right],
	\end{equation}
which indicates a \emph{linearly} growing envelope, modulated by a fast oscillating function, consistent with our exact calculations in Fig. \ref{fig4}(c). Importantly, $\mathcal{D}(\Delta k_x)$ is a real number (related to the residue of the two poles) that becomes larger and larger as we approach the EP (further details in \cite{SM}). These results are a direct consequence of conditions (\ref{1st_cindition})-(\ref{2nd_cindition}), and indicate that, as we approach the EP from higher frequencies, $ \omega \rightarrow \omega_{\mathrm{EP}}^+ $, two oscillatory modes with constant amplitude tend to coalesce into a single linearly-growing mode -- also known as a Jordan mode \cite{note2,Longhi,Graefe,Aaron}. Clearly, such a linear growth should not be interpreted as a form of field amplification experienced by the wave as it propagates along the structure (the wave feels no net gain, and the poles associated with the eigenmodes are still purely real). Our analysis indeed reveals that the linear envelope should be interpreted as the result of a destructive interference pattern between waves with larger and larger amplitude, and closer and closer wavenumber \cite{SM}.


\emph{Topologically-protected loss-immune modal transitions} --- The phase transition through an EP may also be controlled by varying the thickness of the spacing layer between the two gyrotropic media. Fig. \ref{fig3}(d) shows the guided-wave pole constellation, in the complex wavenumber plane, for different gap thicknesses. Considering the same parameters as in Fig. \ref{fig3}(a), and  $ \delta = -\delta = 0.02173 \omega_p $, one can find an EP for $ h = 0.35 \lambda_0 $, where $ \lambda_0 $ is the free-space wavelength at the source frequency, which has been set at $ \omega_0 = \omega_{\mathrm{EP}} = \omega_p / 0.904  $. When $ h > 0.35 \lambda_0 $ the poles form a complex conjugate pair; at $ h= 0.35 \lambda_0 $ the two poles collide on the real axis (EP); then, as $ h \rightarrow 0 $, the EP unfolds into two different poles moving away from each other on the real axis: one tends to $+ \infty $, while the other stops at a finite positive value. 

\begin{figure}[bh!]
	\begin{center}
		\noindent \includegraphics[width=0.475\textwidth]{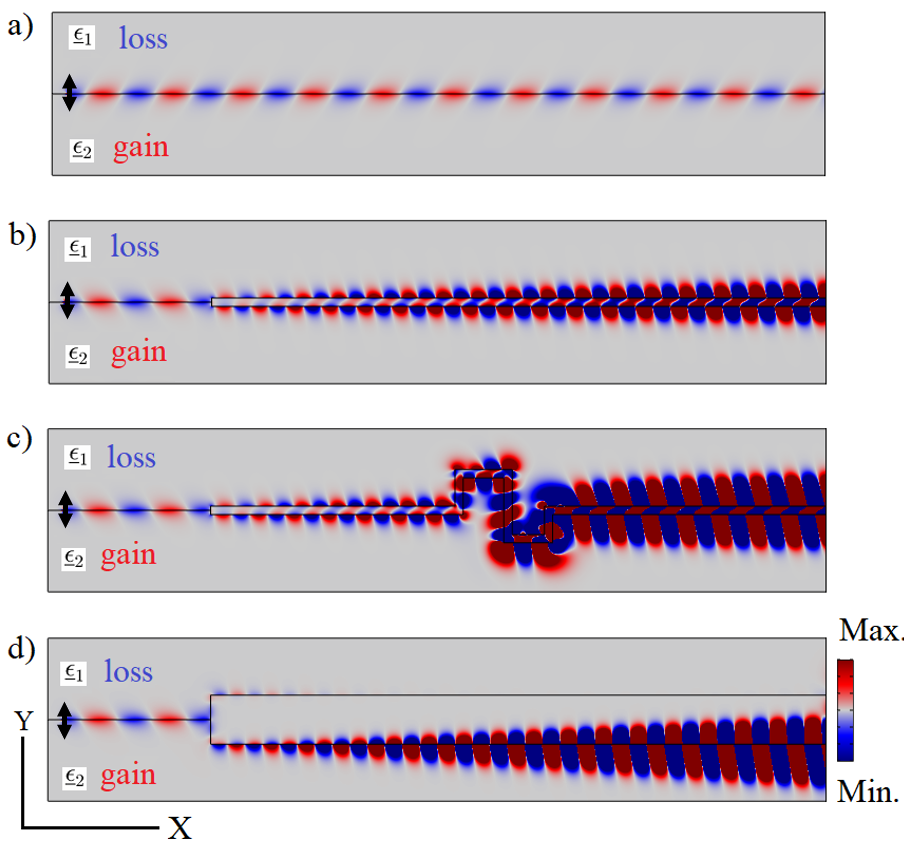}
	\end{center}
	\caption{Magnetic field distributions (time-snapshots) in topological non-Hermitian wave-guiding structures, excited by a dipolar source (black arrow). Different configurations are considered: (a) no gap between the gyrotropic layers (poles on the real axis); (b) a gap of thickness $ h=0.35 \lambda_0 $ is introduced (poles merge at an EP); (c) same as (b) but with a large defect along the waveguide; (d) gap of thickness $ h > 0.35 \lambda_0$ (complex-conjugate poles). Corresponding time-harmonic animations are available in \cite{SM}.}
	\label{fig5}
\end{figure}

To study different waveguide configurations, corresponding to different scenarios in Fig. \ref{fig3}(d), we performed full-wave numerical simulations using a commercial finite-element software \cite{COMSOL}. In Fig. \ref{fig5}(a), we show the steady-state field distribution for a TNT waveguide with spacer thickness $h = 0$, excited by a point source. As expected, a unidirectional surface wave propagates, with constant amplitude, along the interface. If we then introduce a step in the wave-guiding structure, opening an opaque gap of thickness $ h= 0.35 \lambda_0 $ between the two gyrotropic layers (Fig. \ref{fig5}(b)), the incident wave immediately ``jumps'' to the EP regime upon encountering the step, without any reflection because of the topological nature of the involved modes. Rather remarkably, this \emph{topologically-protected modal transition} directly transforms an incident surface wave into a linearly growing (Jordan) mode operating very close to an EP of the parameter space. Even more interesting, the wave continues propagating without any back-scattering even in the presence of large defects (Fig. \ref{fig5}(c)), and grows in amplitude even in the lossy region, in drastic contrast with any conventional wave-guiding structure. For comparison, we show in Fig. \ref{fig5}(d) what happens if the step in the waveguide is larger, bringing the system into the regime beyond the EP, in which conventional exponentially decaying and growing waves are observed.

\emph{Conclusion} --- In summary, we have theoretically demonstrated, for the first time, a non-Hermitian wave-guiding structure that exhibits an \emph{exceptional point between topologically-protected modes}. 
By operating near this degeneracy, we can realize unidirectional modes with low group velocity and inherent immunity to losses and to back-scattering. This topologically-protected behavior would allow rapid, non-adiabatic transitions within the parameter space $\Omega=(\omega,k_x,h,\delta)$ of the wave-guiding structure, without any back-scattering, as long as the initial and final states are within the bulk-modes band gap (hence they have the same gap Chern number). For example, a waveguide operating far from the exceptional point $\Omega_{EP}$ could be excited by a given temporal signal without inducing any instabilities, and then, by changing the parameters in time and/or space, the waveguide may be brought close to $\Omega_{EP}$ to take advantage of the ultra-low values of group velocity in that regime. These exciting possibilities and the temporal analysis of topological non-Hermitian waveguides will be the subject of future works. 

We expect that our theoretical predictions may be experimentally tested, for example, using adjacent oppositely-biased graphene sheets, where the optical gain in one of the sheets may be obtained by pumping graphene edge plasmons by a direct current, in which case the plasmons may gain energy from the kinetic energy of the current carriers, as proposed, e.g., in \cite{Graphene_gain}. 

We believe that our theoretical findings may open new uncharted directions in the centuries-old problem of guiding waves with arbitrary flexibility and robustness against losses and scattering. On a more general note, these results show the great potential of combining ideas of topological and non-Hermitian photonics.


\begin{thebibliography}{99}
\bibitem{note1} A configuration with unequal loss and gain in the system never leads to an EP on the real axis because in the imbalanced condition one of the poles departs from the wave number complex plane real axis earlier than the other and therefore they never coalesce to form a second order rood of the dispersion equation.
\bibitem{note2} A linear mode of this type is also known as a Jordan mode because of the appearance of Jordan blocks in the effective Hamiltonian of the system, associated with the degenerate eigenvalues that merge at the EP \cite{Longhi,Aaron,Graefe}
\end{thebibliography}
\end{document}